\documentclass[12pt]{JHEP3}
\usepackage{bm}
\usepackage{amssymb,amsmath,amsthm}
\usepackage{mathrsfs} 
\usepackage{marginnote}
\RequirePackage{graphicx}

\newcommand{\be}{\begin{equation}} 
\newcommand{\ee}{\end{equation}}
\newcommand{\bea}{\begin{eqnarray}}
\newcommand{\eea}{\end{eqnarray}}

\newcommand{\gapp}{\mathrel{\raise.3ex\hbox{$>$}\mkern-14mu
              \lower0.6ex\hbox{$\sim$}}}
\newcommand{\lapp}{\mathrel{\raise.3ex\hbox{$<$}\mkern-14mu
              \lower0.6ex\hbox{$\sim$}}}

\newcommand\lsim{\lesssim}
\newcommand\gsim{\gtrsim}

\newcommand\vev[1]{{\langle {#1} \rangle}}
\renewcommand\({\left(}
\renewcommand\){\right)}
\renewcommand\[{\left[}
\renewcommand\]{\right]}

\newcommand\eq[1]{Eq.~(\ref{#1})}
\newcommand\eqs[2]{Eqs.~(\ref{#1}) and (\ref{#2})}
\newcommand\eqss[3]{Eqs.~(\ref{#1}), (\ref{#2}), and (\ref{#3})}
\newcommand\eqsss[4]{Eqs.~(\ref{#1}), (\ref{#2}), (\ref{#3})
and (\ref{#4})}

\newcommand\eqreff[1]{(\ref{#1})}

\newcommand\pa{\partial}

\newcommand\mpl{M_{\rm P}}

\newcommand{\dlabel}[1]{\label{#1}}

\def\calp{{\cal P}}

\def\calpz{{\calp_\zeta}}

\newcommand\bfa{{\mathbf A}}

\newcommand\bfk{{\mathbf k}}

\newcommand\bfw{{\mathbf W}}
\newcommand\bfx{{\mathbf x}}

\newcommand\GeV{\,\mbox{GeV}}

\newcommand\sub[1]{_{\rm #1}}
\newcommand\su[1]{^{\rm #1}}

\newcommand\mone{^{-1}}
\newcommand\mtwo{^{-2}}
\newcommand\mthree{^{-3}}

\newcommand\mfive{^{-5}}

\newcommand\half{^{1/2}}

\newcommand\quarter{^{1/4}}

\newcommand\threehalf{^{3/2}}

\newcommand\mn{{\mu\nu}}

\newcommand{\fnl}{f\sub{NL}}

\newcommand\bfkp{{{\bfk}'}}

\newcommand{\calpzphi}{\calp_{\zeta_\phi}}

\newcommand{\zetaphi}{\zeta_\phi}

\newcommand{\phiw}{\phi \sub w}
\newcommand{\hatphiw}{\hat \phi \sub w}
\newcommand{\zetaw}{\zeta  \sub w}
\newcommand{\rhow}{\rho \sub w}

\newcommand{\tw}{t \sub w}
\newcommand{\tf}{t \sub f}
\newcommand{\zetap}{{\zeta_+}}

\title {Statistically anisotropic curvature perturbation generated during the waterfall}

\author{David H. Lyth \\
Consortium for Fundamental Physics, Cosmology and Astroparticle Group, 
Department of Physics, Lancaster University, 
Lancaster LA1 4YB, UK \\ 
 E-mail: \email{d.lyth@lancaster.ac.uk}}
\author{Mindaugas Kar\v{c}iauskas \\ 
CAFPE and Departamento de F\'isica Te\'orica y del Cosmos, Universidad
de Granada, Granada-18071, Spain \\ 
E-mail: \email{mindaugas@ugr.es}}

\abstract{If the waterfall 
field of hybrid inflation  couples to a $U(1)$ gauge field,  the waterfall
can generate a 
statistically anisotropic contribution to the curvature perturbation. We investigate this
possibility, generalising in several directions the seminal work of Yokoyama and Soda.
The   statistical anisotropy of the bispectrum 
 could be detectable by PLANCK even if the statistical anisotropy 
of the spectrum is too small to detect.}

\preprint{}

\begin{document}

\section{Introduction}

Although there 
is so far no evidence for 
statistical  anisotropy of the primordial curvature perturbation $\zeta$, 
mechanisms have been proposed   for generating it.
 Most of them  invoke a  vector field.

One mechanism takes  the vector field to be  homogeneous during inflation, 
but causes significant anisotropy in the  expansion \cite{jiro1} (for a recent review of this approach see \cite{jirorev}).  Then the perturbations of scalar fields generated from the vacuum fluctuation will be statistically anisotropic, and so too will be $\zeta$  on the usual assumption that it originates from one or more of these perturbations.

We here  invoke a different mechanism
 \cite{jiro,ours}
(for the most recent paper on this approach see \cite{km})\footnote
{The use of a vector field to generate a contribution to $\zeta$ was first mooted in
\cite{kostas}.} 
Taking  the inflationary expansion to be  practically isotropic, this mechanism
generates a  perturbation of the  vector field  from the vacuum
fluctuation, which in turn generates a contribution to $\zeta$.
Results using this mechanism are at best  approximate,  
because the unperturbed part of the vector
field will cause at some level anisotropic expansion which generates 
additional statistical anisotropy
through the first mechanism, but existing calculations ignore that effect and we will do the
same.

We work with the setup of \cite{jiro}.
The vector
field is a $U(1)$ gauge field coupled to the waterfall field of hybrid inflation.\footnote
{The study of this setup with non-Abelian gauge fields can be found in \cite{mind}.} 
The dominant contribution to $\zeta$ is supposed to  come  from the perturbation
of the inflaton field.  But the perturbation of the gauge field
is supposed to generate an additional 
 contribution   during the waterfall that ends inflation.
The  waterfall is taken to be practically instantaneous.
We extend the original treatment of the scenario in several respects.
First, we do not assume  the inflaton potential $V=V_0+m_\phi^2\phi^2/2$ (which is 
ruled out by observation).
 Second, we do not assume that the perturbation of the gauge
field is exactly scale-invariant. Third, we take into account the time-dependence of
the gauge field.

We  take for granted the main ideas of modern cosmology described for
instance in \cite{book}, and use  the  notation and definitions of \cite{ours,book}.
The unperturbed universe has the line element
\be
ds^2 = -dt^2 + a^2(t) \delta_{ij} dx^i dx^j.
\dlabel{ds2}
\ee
For any smooth rotationally invariant quantity
$g(\bfx,t)$,   uniquely defined during some era,
 we can choose a slicing
(fixed $t$) and threading (fixed $\bfx$) and then write
\be
g(\bfx,t) = g(t) + \delta g(\bfx,t)
. \ee
Going to a different slicing with a time displacement
$\delta t(\bfx,t)$, we have  to first order 
\be
[\widetilde {\delta g}(\bfx,\tilde t)] - \delta g(\bfx,t) = 
g(\bfx,t) - g(\bfx,\tilde t) \simeq -\dot g(t) \delta t(\bfx,t)
. \ee
We will invoke this `gauge transformation'  without comment.
In most cases  $g$ is homogeneous on one of the slicings.

\section{The curvature perturbation $\zeta$}

\subsection{Definition and $\delta N$ formula}

To define $\zeta$ one smoothes the metric on a super-horizon scale, and adopts the 
comoving threading and the slicing of uniform energy density $\rho$. Then
\cite{deltan}
\be
\zeta(\bfx,t)\equiv \delta [\ln a(\bfx,t)]
=\delta [\ln \left( a(\bfx,t)/a(t) \right)] \equiv \delta N(\bfx,t)
, \dlabel{deln} \ee
where $a(\bfx,t)$ is the locally defined scale factor (such that a comoving volume element
is proportional to $a^3(\bfx,t)$) and $a(t)$ is its unperturbed value. The number of $e$-folds
of expansion $N(\bfx,t,t_*)$ starts from a slice at time $t_*$
on which $a$ is unperturbed (`flat slice') and ends
on a uniform $\rho$  slice at time $t$. Since the expansion between two flat slices is uniform,
$\delta N$ is independent of $t_*$.
The change in $\zeta$ between two times is
\be
\zeta(\bfx,t_2) - \zeta(\bfx,t_1) = \delta N(\bfx,t_1,t_2)
, \dlabel{deln12} \ee
where now  both the initial and final slices have uniform $\rho$.

By virtue of the smoothing, the energy conservation equation is valid locally:
\be
\dot\rho(t) = 3 \frac{\pa a(\bfx,t)}{\pa t} \( \rho(t) + P(\bfx,t) \)
. \ee
In consequence, $\dot\zeta=0$ during an era when $P(\rho)$ is a unique function.
The success of the BBN calculation shows that $P=\rho/3$ to high accuracy 
 when cosmological scales start to enter the horizon. Then  $\zeta$ has a time-independent
value $\zeta(\bfx)$ that 
is strongly constrained by observation. Within observational errors it is gaussian and
statistically isotropic. Its spectrum is nearly
independent of $k$, with 
\be
\calpz(k) \simeq   (5\times 10\mfive)^2
.  \dlabel{calpzobs}\ee
For the reduced bispectrum \cite{komatsu}
$\fnl$, current observation give $|\fnl|\lsim 100$ and barring a 
detection PLANCK will give $|\fnl|\lsim 10$.  For $\fnl$  to ever be observable we need
$|\fnl|\gsim 1$. 

We will work to first order in $\zeta$, so that
\be
\zeta(\bfx,t) = H(t) \delta t_{{\rm f}\rho}
, \dlabel{delt} \ee
where $\delta t_{{\rm f}\rho}$ is the time displacement from the flat slice to the 
the  uniform-$\rho$ slice. 
A  second-order calculation of  $\zeta$ is   needed only to treat very small
non-gaussianity corresponding to $|\fnl|\lsim 1$.

To explain the near scale-invariance of the observed $\calpz(k)$,  it is usually supposed
that $N(\bfx,t)$ is determined by the values of one or more fields $\phi_i(\bfx,t)$,
evaluated during inflation at an epoch $t_*$ when relevant 
scales have left the horizon: 
\be
N(\bfx,t) = N(\phi_1^*(\bfx),\phi_2^*(\bfx),\cdots,t)
. \ee
The fields are defined on a flat slice and denoting their values by $\phi_i^*$
we write
\be
\phi_i^*(\bfx) = \phi_i^* + \delta \phi_i^*(\bfx)
 \ee
and  \cite{lr}
\be
\zeta(\bfx,t) = \sum N_i  \delta\phi_i^*(\bfx) + \frac12 \sum_{ij} 
N_{ij}
\delta\phi_i^*(\bfx) \delta\phi_j^*(\bfx) + \cdots,
\dlabel{deltan} \ee
where  a subscript $i$ denotes $\pa/\phi_i^*$ evaluated at $\phi_i^*(\bfx)=\phi_i^*$.
The $\phi_i$ are usually taken to be scalar fields, but it has been proposed
\cite{jiro,ours} that some or all of them may be components of a vector field.

On each scale $k$, the 
field perturbations are generated from the vacuum fluctuation at horizon exit.
Ignoring scales leaving the horizon after $t_*$ 
 \eq{deltan} defines
a classical quantity $\zeta$, which is independent of the choice of $t_*$.
In general it depends on $t$, settling down to the observed quantity $\zeta(\bfx)$  by
some time $t\sub f$. 

Cosmological scales have a fairly narrow range $\Delta k\sim 15$ or so.
Choosing $t_*$ as the epoch when the shortest scale leaves the horizon,
scalar fields with the canonical kinetic term, that satisfy  the 
slow-roll approximation, have a nearly Gaussian uncorrelated 
perturbations  with spectrum 
$\calp_{\delta\phi_i*}\simeq (H/2\pi)^2$.
To have the observed nearly gaussian
$\zeta(\bfx)$ \eq{deltan} has to be dominated by one or more linear terms (at least
when $t=\tf$).
Keeping only linear terms,
\be
\calpz(\bfx,t) \simeq \sum N_i^2 \calp_{\delta\phi_i^*} + \cdots
, \dlabel{deltan2}\ee
where the terms exhibited correspond to scalar fields, and the dots indicate
vector field contributions. The contribution of the latter is positive
like the rest \cite{ours}.
 The non-linear terms may  give non-gaussianity that can be observed
in the future.

\subsection{Slow-roll inflation}

\dlabel{2.2} 

Inflation corresponds 
to $\epsilon_H< 1$ where $\epsilon_H \equiv -\dot H/ H^2$.
During inflation each 
coordinate wavenumber $k$ (scale) leaves the horizon when $k=aH$, and
we  are interested only in the 
era of inflation after horizon exit for the biggest observable
scale  $k\sim a_0H_0$. 
Here $H\equiv \dot a(t)/a(t)$, $k$ is the coordinate wavenumber and
the subscript 0 denotes the present. 
We need $\epsilon_H\ll 1$ at least while cosmological scales leave the horizon
to generate the nearly scale-invariant $\calpz(k)$.

We are interested in single-field slow-roll inflation. Here, the only field 
with significant variation during inflation is the inflaton. Its unperturbed value
$\phi(t)$ 
 satisfies  the slow-roll approximation.
\bea
3H\dot \phi  &\simeq& -V'(\phi), \dlabel{phidot}  \\
\epsilon &\equiv&  \frac12\mpl^2 (V'/V)^2\simeq \epsilon_H \ll 1 \\
\rho(t) &=&  3\mpl^2 H^2 \simeq V(\phi). \dlabel{rho}
\eea 

The perturbation $\delta \phi_*$ can be removed by a shift $\delta t_*(\bfx,t)$,
which means that it generates a time-independent contribution to 
$\zeta_\phi$. Since the $\fnl$ generated by $\zeta_\phi$ is
negligible \cite{maldacena}, we can work to 
 first order in $\delta\phi_*$,
\be
\zetaphi(\bfx) = - (H/\dot\phi) \delta\phi_*(\bfx)
. \dlabel{zetaphi} \ee

Choosing $t_*$ as the epoch of horizon exit, we find the spectral index is given by
\bea
\calpzphi(k) &\simeq & \frac1{2\epsilon \mpl^2} \( \frac H{2\pi} \)^2
  \dlabel{calpzphi} \\
n_\phi(k) -1 &\equiv&  d\calpzphi/d\ln k = 2\eta - 6 \epsilon, 
\eea
where $\eta\equiv \mpl^2 V''/V$ with $|\eta|\ll 1$, and the 
 right hand sides are  evaluated at the  epoch of horizon exit $aH=k$.

Although it is not our central concern, we mention at this point the case
of multi-field slow-roll inflation, where 
two or more fields vary significantly. Taking $\phi$ to be the field pointing
along the trajectory at horizon exit,
\eq{calpzphi} still applies
to that case.

If $\zeta$ depends only on the part of the action that we are considering,
$\zetaphi(k)$ can be identified with the observed quantity $\zeta(k)$.
More generally we have
\be
\calpzphi(k) \lsim \calp(k) \simeq (5\times 10\mfive)^2.
\dlabel{calpzphibound}
\ee

This inequality is important for two reasons.
First, it makes the tensor fraction
  $r\leq 16\epsilon$.
Then the  slow roll approximation gives 
what has been called the Lyth bound, on the variation
$\Delta \phi$ of the inflaton field after the observable universe leaves the horizon.
Without any assumption about the function $\epsilon(\phi)$ after the first few $e$-folds,
one finds \cite{mybound}
\be
10\mone   \(\Delta \phi\ / \mpl \)^2 \gsim 16\epsilon   \geq  r. 
\ee
If $\epsilon(\phi(t))$ increases with time, $10\mone$
 in the above expression is replaced
\cite{hilltop} by $0.0003$. According to these results, an observable $r$ cannot be
obtained with $\Delta \phi \ll \mpl$ (small-field model).

The other use of the inequality is for curvaton-type models, where
 $\zeta$ is generated almost entirely by the 
perturbation of some field  that has a negligible effect during
inflation. Then   $\calpzphi$ will be
 negligible compared with $\calpz$, and 
 $r$ will be negligible compared with $16\epsilon$ so that it is unobservable.

All of this assumes slow-roll inflation, in which it is assumed
that there is no time-dependent field 
 except the slowly-rolling inflaton fields.
If there is such a field 
 the shift in the initial time generated by  $\delta\phi_*$ will be 
accompanied by a shift in the value of that field, which could allow
$\zetaphi$ to be time-dependent and 
avoid the inequality \eqreff{calpzphibound}.

The possibility of avoiding this inequality was mooted in \cite{bkr,martin}
but they did not find a mechanism. One can easily avoid the inequality
by abandoning the canonical kinetic term for the inflaton
\cite{slava} and we are for the
first time pointing 
to a possible mechanism with the canonical kinetic term.

\section{The model}

\subsection{Hybrid inflation}

\dlabel{earlier}

 The relevant part of the action is
\be
S = \int d^4x \sqrt{-g} \[ \frac12 \mpl^2 R
- \frac12 \pa_\mu \phi \pa^\mu \phi - \frac12 \pa_\mu \chi \pa^\mu \chi 
 - \frac14 f^2(\phi) F_\mn F^\mn  - V \]
, \dlabel{action} \ee
with $F_\mn \equiv \pa_\mu B_\nu - \pa_\nu B_\mu$ and  
 $B_\mu$  a $U(1)$ gauge field.
Following \cite{ours} we 
 use  the gauge with $B_0=\pa_i B_i=0$, and work with $A_i\equiv
B_i/a$ which is the field defined with respect to the locally orthonormal basis
(as opposed to $B_i$ which is defined with respect to the coordinate basis).
The raised component is $A^i=A_i$ (as opposed to  $B^i=B_i/a^2$).
We also define $\bfw\equiv f\bfa$, which would be 
the canonically normalized field if $f$ were constant. To fix the normalization
of $f$, we set $f=1$ at a time $\tw$ just before the waterfall begins.

Before the waterfall the   potential is 
\bea
V(\phi,\chi,A)  &=&  V_0 + V(\phi) 
+\frac12 m^2(\phi,A)  \chi^2 + \frac14\lambda \chi^4 +\frac12 \mu^2 A^2
 \dlabel{fullpot2}  \\
m^2(\phi,A) &\equiv &  h^2  A^2 + g^2\phi^2 - m^2
. \dlabel{msq} \eea
The 
 waterfall field $\chi$ is supposed to be the radial part of a complex field
which is charged under the $U(1)$, generating the first term of \eq{msq}.

This is the usual hybrid inflation potential \cite{andreihybrid,ourhybrid}
except for the presence of $A$.
We assume that the values of the parameters and fields give what has been
called standard hybrid inflation \cite{p11}. 
At each location, the 
 waterfall begins when
 $m^2(\phi(\bfx,t),A(\bfx,t))$ falls to zero.   Before it begins,
the waterfall field
$\chi$ vanishes up to a vacuum fluctuation which is set to zero, 
 and we have slow-roll inflation with
\be
V = V_0 + V(\phi)\simeq V_0 
. \ee
Cosmological scales are supposed
to leave the  horizon before the waterfall begins.
 We will take $H$ to be constant which
is typically a good approximation. 

\subsection{Field equations}

During the waterfall, $\chi$ moves to it's vev and then inflation ends.
We will assume that the duration of the waterfall is so short that it can
be taken to occur on a practically unique slice of spacetime. The evolution
of $\phi$ and $A$  is therefore  required only before the waterfall begins.

To work out the field equations, previous authors have  taken $f(\phi)$
to be a function of time with $f \propto a^\alpha(t)$, and have
taken spacetime to be unperturbed. Then the 
action \eqreff{action} gives for the unperturbed fields
\bea
\ddot \phi(t) + 3H\dot\phi(t) + V'(\phi(t)) = 0 \\
\ddot \bfw(t) + 3H\dot\bfw(t) + \mu^2 \bfw(t) = 0, 
\eea
where
\be
\mu^2 \equiv H^2 (2+\alpha)(1-\alpha)  \dlabel{mu}
. \ee
By virtue of the flatness conditions on the potential ($\epsilon \ll 1$
and $|\eta| \ll 1$), the first expression is expected to give the slow-roll
approximation \eqreff{phidot} more or less independently of the initial
condition. Similarly, the second equation is expected to give the 
slow-roll approximation 
\be
3H\dot\bfw \simeq - \mu^2 \bfw
 \dlabel{bfaeq2} \ee
if $|\mu|^2 \ll H^2 $ which we assume. 

In terms of $W$, the coupling $h^2 A^2\chi^2$ becomes
$\tilde h^2 W^2 \chi^2$, where $\tilde h \equiv hf$. Before  
horizon exit on cosmological  scales
 we are taking $\bfw$ to be a practically free field corresponding to
$\tilde h\ll 1$. With $\alpha\simeq 1$  this would give at $t=\tw$  
a tiny coupling $ h\ll e^{-N_k}$ which would have practically no effect. 
We therefore assume $\alpha \simeq -2$.

For the  first order perturbations, $f\propto a^\alpha$  gives
\bea
\delta \ddot \phi_\bfk(t) + 3H \delta \dot \phi_\bfk(t) + 
((k/a)^2 + V''(\phi(t))\delta\phi_\bfk = 0 \dlabel{ddotdelphi}  \\
\delta \ddot \bfw_\bfk(t) + 3H \delta \dot \bfw_\bfk(t) + 
((k/a)^2 + \mu^2)\delta\bfw_\bfk = 0 \dlabel{ddotdelw}
. \eea

Since $\phi$ and $\bfw$ are slowly varying,
this flat spacetime calculation is  expected to hold
in the perturbed universe on the flat slicing.
 It is expected because
 \cite{book,book1} the effect of the metric perturbation
(back-reaction) on \eq{ddotdelw} is 
proportional to the small quantity $\dot  \bfw(t)$.

Since we are assuming $f(\phi)$, the choice $f\propto a^\alpha$ 
 corresponds to 
\be
f\propto \exp \( \alpha \int^\phi  [\sqrt{2\epsilon(\phi)} \mpl]\mone d\phi \) 
. \dlabel{fofphi}  \ee
This gives the perturbation
\be
\delta f/f = \frac \alpha{\sqrt{2\epsilon}\mpl} \delta\phi
. \dlabel{deltaf} \ee

Since $f$ is a function of $\phi$, the term $\propto f 
F_\mn F^\mn$ in the action couples 
$\phi$ and $\bfw$ so that the right hand sides of 
\eqsss{phidot}{bfaeq2}{ddotdelphi}{ddotdelw}  are nonzero.
We calculate them in the Appendix, and show that they are negligible
if 
\be
\frac{\rho_B}{\epsilon \rho} =\frac12 \frac{\dot W^2}{\epsilon \rho}
\simeq
\frac12\frac{\mu^2W^2}{\epsilon V} \simeq \frac16 
\frac{ W^2 }{\epsilon \mpl^2}
\ll 1
, \dlabel{rhowcon} \ee
where $\rho_W$ is the energy density of $\bfw$. 
We will assume this condition. Note that it implies
$\rho_W \ll \rho$, which is anyway needed because we are taking the 
expansion of the universe to be isotropic. 
From \eq{calpzphibound}, the condition is
guaranteed if $W/H\lsim 10^5$.

\subsection{Spectrum of $\bfw$}

The evolution equation for $\bfw(\bfx,t)$ is the same as that of a 
free scalar field with mass-squared $\mu^2$, and we are assuming
$|\mu|^2 \ll H^2$.
Treating  $\delta W_\bfk$ as  an operator and assuming the vacuum state
well before horizon exit gives the approximately scale-independent 
vacuum expectation value
\bea
\frac{k^3}{2\pi^2} \vev{ \delta W_\bfk^i(t) \delta W_\bfkp^j(t) }
&=& \(\delta^{ij} - \hat k^i \hat k^j \)
\delta^3(\bfk+\bfkp) \( \frac H{2\pi } \)^2
  \( \frac k {a(t)  H }  \)^{n\sub{vec} - 1 } \dlabel{aspec2} \\
~ n\sub{vec}-1   &=& 2\mu^2/3H^2, 
~ \dlabel{nvec}
\eea
where hats denote unit vectors. According to \eq{bfaeq2},  $\delta \bfw_\bfk$ has constant phase which means that
it can be treated as a classical quantity with this correlator.

We are interested in $t=\tw$, when
\be
 \( \frac k {a(\tw)  H }  \)^{n\sub{vec} - 1 } 
= e^{-N_k (n\sub{vec} -1 ) } \equiv e^x
. \ee
Also, we  are interested in cosmological
scales, which  have a range $\Delta N_k \sim 15$ and
a typical central value $N_k\sim 50$.

The decomposition
\be
\bfw(\bfx,t) = \bfw(t) + \delta \bfw(\bfx,t)
\dlabel{deltaW}
\ee
is made in some box of coordinate size $L$
around the observable universe, with $\bfw(t)$
the average within the box.
After smoothing on a cosmological scale $k$, the spatial average of
$(\delta W)^2$ (evaluated within a region not many orders of magnitude
bigger than the observable universe) is of order $\ln (kL) H$ 
and we assume that the box is not too big,
$\ln(kL)$ roughly of order 1.  
Guided by these results, we  assume $W(t) \gg H$,
which is reasonable because $W^2(t)$ at a typical position is expected to be
at least of order the mean square of $(\delta W)^2$ evaluated within
a much larger box \cite{mybox}. 

\section{Including the waterfall contribution}

\subsection{End-of-inflation formula}

At an epoch $t_+$ just after inflation ends, 
\be
\zeta(\bfx,t_+) =\zetap(\bfx)  \equiv \zetaphi(\bfx) + \zetaw(\bfx),
\dlabel{zetap} 
\ee
 where $\zetaw$ is the waterfall
contribution.

To calculate 
$\zetaw$,  we 
suppose that the 
waterfall happens very quickly  
so that it can be regarded as taking place on a single spacetime
slice.
Then   \cite{p11,endinf}
\be
\zetaw(\bfx)= H \delta t_{12}(\bfx) =  H\[ \frac{ \delta\rhow(\bfx) }{\dot\rho(\tw)} - 
\frac{ \delta\rhow(\bfx)}{\dot \rho(t_+) }\] \simeq  H \frac
{ \delta\rhow(\bfx) }{ \dot\rho(\tw) } \simeq H \delta t_{\rho\rm w}
. \dlabel{zwofrho} \ee
In this equation, $\delta t_{12}(\bfx)$ is the proper time elapsing between a uniform-$\rho$
slice at time $\tw$ just before the waterfall, and a uniform-$\rho$ slice at time
$t_+$ just after the waterfall. Because $|\dot\rho|$ is much smaller during inflation than 
afterwards, $\delta t_{12}$ is practically the same  $\delta t_{\rho \rm{w}}$,
the displacement from the initial uniform-$\rho$ slice to the waterfall slice.
Using \eq{delt}, we see that 
\be
\zeta(\bfx,t_+) = H \delta t(\bfx) \dlabel{zofdelt}
, \ee
 where $\delta t$
is the displacement from the flat slice at $\tw$ to the waterfall slice.

This end-of-inflation formula  actually holds if the waterfall slice is replaced by  
any sufficiently brief  transition from inflation to non-inflation.
In \cite{p11} it is invoked for the transition beginning {\em during} the waterfall,
at the epoch when the evolution of $\chi$ becomes non-linear. 
We are here applying  it to the  entire waterfall.
It   was  first given \cite{endinf} with $A$ in \eq{msq} 
replaced by a   scalar field. In \cite{endinf} the slope of the potential
in the $A$ direction was assumed to be negligible corresponding to
single-field hybrid inflation, and the same assumption was made in several
later papers  \cite{endapps1}. The assumption was relaxed in
 \cite{endapps2,endapps3,endapps4}, corresponding to what has been  called \cite{endapps3}  multi-brid inflation. Following \cite{jiro} we are here taking  $A$ to be the magnitude of a $U(1)$ gauge field. One can also  replace $A$ by a non-Abelian gauge field \cite{mind,toni,mind2}.

Before continuing, let us ask what 
is required to make the waterfall sufficiently brief.
Just to have the error $|\Delta \zetaw| \ll |\zetaw|$, one presumably
needs 
 $\Delta t(\bfx) \ll |\delta t_{12}(\bfx)| $, 
which is equivalent to   \cite{p11}.\footnote
{The second inequality allows for contributions to $\zeta(\bfx)$ that might be
generated after inflation by fields different from $\phi$ and $A$, and assumes that 
the contribution of the latter undergoes no further change.}
\be
H\Delta t \ll \calp_{\zetaw}\half \leq \calpz\half
 =  5\times 10\mfive 
. \dlabel{deltat} \ee
During the waterfall $m^2(\phi,A)$ goes from $0$ to 
$-m^2$ of the waterfall, with $m^2\gsim   H^2$. 
 We therefore expect
$\Delta t$ to  be at least of order $1/m$, and 
we need $m/H \ll \sqrt {\mpl/H}$ so that
$\lambda \ll 1$ \cite{p11}. 
Hence \eq{deltat} requires an inflation scale $H/\mpl \ll 10^9\GeV$
or $V\quarter \ll 10^{14}\GeV$. This rules out GUT hybrid inflation 
$(V\quarter \sim 10^{15}$ or so) but easily allows inflation at the scale of supersymmetry
breaking ($H$ of order the gravitino mass $\lsim 10^5\GeV$ or so).

A stronger requirement might be needed to justify a  
calculation of the non-gaussianity parameter
$\fnl$, because it refers to the non-gaussian part of $\zeta$ that is only of order
$\calpz\half \fnl \lsim 10\mthree$. A reasonable estimate for
 $|\Delta \zetaw/\zetaw|$  might be  $H\Delta t/\calp_{\zetaw}\half$. Then, in the worst
case that $\Delta \zetaw$ is completely non-gaussian, one  would
 require the  very low inflation scale
$H/\mpl \ll  10^{-18} f\sub{NL}^2$.
We proceed on the assumption that  the inflation scale is sufficiently small.

\subsection{Waterfall contribution in our model}

\dlabel{3.2}

Without at first specifying the nature of $A$, we now  calculate $\delta t$. 
The fields on the waterfall slice have values given by 
$m^2(\phiw(\bfx),A\sub w(\bfx)) = 0$. It was noted in \cite{mind} that the time dependence of a waterfall slice could be important. Thus let us define a `time-dependent waterfall slice' $\phiw(\bfx,t)$ by
\be
m^2(\phiw(\bfx,t),A(\bfx,t)) = 0. 
\dlabel{phiw} 
\ee
(If this equation has more than one solution $\phiw(\bfx,t)$ we choose one of them.)
If   $\dot\phi\sub w$ is  negligible,
   the waterfall occurs when $\phi(\bfx,t)$
falls to the practically time-independent waterfall slice  $\phiw(\bfx)$. 
If instead $\dot\phi$ is negligible, the waterfall occurs  when the time-dependent
waterfall slice $\phiw(\bfx,t)$  meets the practically
time-independent field value $\phi(\bfx)$.
Since we deal with hybrid inflation $\dot\phi$ is negative, but $\dot\phiw$ might have
either sign. If it is also negative 
 we need $|\dot \phi\sub w|< |\dot\phi|$
or inflation will never end.

If $\delta t(\bfx)$ is the displacement 
from the flat slice at $\tw$ to the waterfall 
slice, this gives to first order in $\delta t$
\bea
\phi(\bfx,\tw+\delta t(\bfx)) 
&=&\phi(\tw) + \delta \phi(\bfx,\tw) + \dot\phi(\tw) \delta t(\bfx) \\
\phiw(\bfx,\tw+\delta t(\bfx)) 
&=&\phiw(\tw) + \delta \phiw(\bfx,\tw) + \dot\phi\sub w(\tw) \delta t(\bfx)
, \eea
where $\delta\phiw$ is defined on the flat slice.
Setting $\delta t=0$ gives 
$\phi(\tw)=\phiw(\tw)$. Evaluating $\delta t$ we have
\be  
\zeta(\bfx,t_+) = H\delta t(\bfx) = H \frac{\delta \phiw(\bfx,\tw)
-\delta \phi(\bfx,\tw) }{ \dot \phi(\tw) - \dot\phi\sub w(\tw) }
. \dlabel{zeta} \ee

Now we invoke \eq{msq}. 
Discounting  the  strong cancellation $m^2 \simeq h^2 A^2$ it gives
\be
~ \phiw(\bfx,t)  = \frac1g (m^2- h^2 A^2 (\bfx,t) )\half 
\simeq  \frac mg - \frac12 \frac{h^2 A^2(\bfx,t) }{mg} ~
. \dlabel{phiw2} \ee
Using \eqsss{zetaphi}{bfaeq2}{deltaf}{phiw2}  we get\footnote
{Terms involving a product of $\delta\phi$ with itself or
$\delta\bfw$ are dropped because they are negligible.}
\be
\zeta(\bfx,t_+) = \zetaphi(\bfx) \( 1 + 
\frac{\mu^2}{H^2}\frac{2XA^2}{1+2X A^2} \)
+\frac{ \widehat \zeta_w }{1+2XA^2 }
, \dlabel{main} \ee
where $\zetaphi(\bfx)$ is defined by \eq{zetaphi} and
\bea 
\widehat \zeta\sub w &=& 
- X \( \bfw(\tw)  \cdot \delta\bfw(\bfx,\tw) + (\delta W(\bfx,\tw))^2 \) 
\dlabel{zeta7} \\
X&\equiv& h^2/\sqrt{2\epsilon}\mpl mg
. \eea
Previous authors except \cite{hassan} 
ignored the time-dependence of $A$, which means that
they implicitely set  $\dot\phi\sub w(\tw)=0$ to obtain
$\zetaw=\widehat \zeta\sub w$.\footnote
{The  spectrum of the waterfall contribution
 found in  \cite{hassan} is negligible
(smaller than ours by a factor $(1+2XA^2)^2 e^{-2N_k}$). We have not been
able to follow this calculation, which is not
 from first principles because $A$ is treated as a scalar.}

Taking $t_*=\tw$, the second term of \eq{main} is the contribution
of $\delta W_*$ to $\zeta$, which means that the first term 
is the contribution of $\delta\phi_*$. It differs slightly from the 
result found earlier in \eq{zetaphi},  but the difference is not significant;
to  calculate the contribution of $\delta\phi_*$ 
taking  account of $\bfw$, one would have to include
the anisotropy of the expansion of the universe  caused by $\bfw$ which
is beyond the scope of our investigation.  Therefore, at the level
of our calculation there is no change in the usual assumption that
$\zetaphi(\bfx,t)$ is constant.

The formalism that we have given involves  $\delta\phiw$, which is
a function of $\delta\bfa$ and hence of 
both $\delta\phi$ and $\delta\bfw$.
A more direct approach is to use  $\hatphiw(t)$ defined by
$m^2(\hatphiw,(W/f(\hatphiw))^2)$. Then $\delta\hatphiw$
is a function only of $\bfw$,  and 
\be
\dot{\hat\phi}\sub w / \dot\phi = -
\frac{\mu^2}{H^2}\frac{2XA^2}{1+2X A^2} 
, \ee
leading directly to \eq{main}.

\subsection{Anisotropic spectrum and bispectrum}
 Since  $W \gg  H$,  
the linear  term of \eq{zeta7} dominates,
 leading to  \cite{jiro,ours}
\bea
\calpz(k) &=& 
\calp_\zeta\su{iso} \[ 1 -  \beta \( \hat \bfa \cdot \hat \bfk \)^2 \] \\
\calp_\zeta\su{iso} &=& \frac{\calpzphi (1+\beta) }{1+2XH^2 }
, \eea
where
\be
\beta = \frac{h^4 A^2(\tw)}{m^2 g^2} e^x 
. \dlabel{betapred} \ee
Current observation  requires $\beta\lsim 10\mone$, 
and barring a detection PLANCK
will give $\beta \lsim 10\mtwo$ \cite{anis}. 
Using  \eqs{calpzobs}{calpzphi}, the 
 observed value of $\calpz$  requires
\be
 \frac1{2\epsilon \mpl^2} \( \frac H{2\pi} \)^2 \simeq  (5\times 10\mfive)^2
\( 1+2XA^2 \)^2
. \dlabel{constraint} \ee

Including the second term of \eq{zeta7} we find  \cite{jiro,ours}
\be
\fnl = f\sub{NL}\su{iso} \( 1 + 
f\su{ani}(\bfk_1,\bfk_2,\bfk_3) \)
\ee
where   
\be
f\su{ani} = \frac{
 - (\hat \bfa \cdot \hat \bfk_1)^2
-(\hat \bfa \cdot \hat \bfk_2)^2 + (\hat \bfk_1 \cdot \hat \bfk_2)
(\hat \bfa \cdot \hat \bfk_1)(\hat \bfa \cdot \hat \bfk_2)  }{
\sum k_i^3/ k_3^3  }
+ \mbox{ 2 perms.}
\ee 
(with $\bfk_1+\bfk_2+\bfk_3=0$)
and 
\be
\frac65 f\sub{NL}\su{iso} = \frac{1+2XA^2}{ XA^2 } \beta^2
\simeq 3 \( 100 \beta \)\threehalf
\frac H { A(\tw) } \frac{1+2XA^2}{XA^2}  e^{x/2} 
.\dlabel{fnlpred} \ee
(For the final expression we used \eqss{calpzphi}{betapred}{constraint}.)
The last two factors, omitted in the original calculation \cite{jiro}, allow
 PLANCK to   detect  $f\sub{NL}\su{iso}$  even if it does
not detect $\beta$.

This  calculation ignores the  
 time-dependence of
$\epsilon$. 
Allowing time-dependence for $\epsilon$ would multiply $\zetaw$ by a factor
$[\epsilon(t\sub{cos})/\epsilon(\tw) ]\half$,
where $t\sub{cos}$ is 
the epoch of horizon exit for a  typical cosmological scale.
The factor might be significantly different from 1, 
but there is little point in including
it because its effect is indistinguishable
from the effect of the tilt factor $e^x$.

\section{Conclusion}

If the waterfall of hybrid inflation is sufficiently brief, it takes place on a practically unique slice of spacetime. Then the waterfall slice contributes to the curvature perturbation $\zeta$, if its location depends on some field $A$ different from both the inflaton and the waterfall field. 

We generalised  in several directions the  model of
Yokoyama and Soda \cite{jiro}, that takes  $A$ to be a $U(1)$ gauge field.
The model  makes $\zeta$ statistically anisotropic, and we 
 find that 
the  prediction for  $\fnl$  
could be verified by PLANCK, even if the prediction 
for the anisotropy of $\calpz$ is too small to be detected.

The  weak point of the model is the special form \eq{fofphi}, that is required
to get the gauge kinetic function $f(\phi)\propto a\mtwo$. 
We are not aware of any well-motivated hybrid 
inflaton potential that would lead to a well-motivated $f(\phi)$.
 This is in contrast  with the case of non-hybrid inflation
\cite{gaugefn}, where one can take
$f(\phi)$ and $V(\phi)$  to have 
exponentially increasing behaviour that might be reasonable in string theory
\cite{exponent},
and which could correspond to an  attractor (late-time limit) 
\cite{attract}. 

In the course of our investigation we noticed that the presence, during 
slow-roll inflation, of a 
time-dependent field different from the inflaton  might allow a significant
decrease in  the spectrum
of the curvature perturbation after horizon exit. That does not however
happen  in our case.

\section{Acknowledgments}
DHL 
 acknowledges support from the Lancaster-Manchester-Sheffield Consortium for
Fundamental Physics under STFC grant ST/J00418/1, and from
 UNILHC23792, European Research and Training Network (RTN) grant. 
MK is supported by the grants CPAN CSD2007-00042
and MICINN (FIS2010-17395).
We thank  K.~Dimopoulos, H.~Firouzjahi, 
J.~Soda and  S.~Yokoyama for discussion in the early stage of this work. 

\appendix
\section{Equations of Motion for $\phi(\bfx,t)$ and $\mathbf W(\bfx,t)$}

Extremizing the action in \eq{action} with respect to fields $\phi$, $B_\mu$ and their derivatives we obtain field equations 
\begin{eqnarray}
\left[\partial_{\mu}+\partial_{\mu}\ln\sqrt{-g}\right]\partial^{\mu}\phi+V'(\phi)+\frac{1}{2}ff'(\phi)F_{\mu\nu}F^{\mu\nu}&=&0;\\
\left[\partial_{\mu}+\partial_{\mu}\ln\sqrt{-g}\right]fF^{\mu\nu}&=&0,
\end{eqnarray}
where $g\equiv \mathrm{det} (g_{\mu\nu})$ and $f_{,\phi}\equiv\partial f/\partial\phi$. Choosing the temporal gauge $B_0 = 0$ and a line element of the unperturbed universe in \eq{ds2}, one finds equations of motion for the fields $\phi(\bfx,t)$ and $\mathbf B(\bfx,t)$
\begin{eqnarray}
\ddot{\phi}+3H\dot{\phi}-a^{-2}\nabla^{2}\phi+V'(\phi) & = & -\frac{1}{2}f(\phi)f'(\phi)F_{\mu\nu}F^{\mu\nu},\\ \label{EoM-phi}
\ddot{B}_{i}+\left(H+2\frac{\dot f}{f}\right)\dot{B}_{i}-a^{-2}\nabla^{2}B_{i} & = & a^{-2}2\frac{\partial_{j}f}{f}\partial_{j}B_{i}, \label{EoM-W}
\end{eqnarray}
Recasting the above equations in terms of $\mathbf W \equiv f \mathbf B /a$ and dropping gradient terms, one arrives at equations of motion for homogeneous fields $\phi(t)$ and $\bfw (t)$
\begin{equation}
\ddot{\phi}+3H\dot{\phi}+V'=\frac{f'(\phi)}{f}\left[\dot{\mathbf{W}}+\left(H-\frac{\dot{f}}{f}\right)\mathbf{W}\right]^{2}, \dlabel{hom-phi}
\ee
\be
\ddot{\mathbf{W}}+3H\dot{\mathbf{W}}+\left(2H^{2}-H\frac{\dot{f}}{f}-\frac{\ddot{f}}{f}\right)\mathbf{W}=0, \dlabel{hom-W}
\end{equation}
where we also used $\dot H \simeq 0$.

Decomposing the field $\bfw(\bfx,t)$ as in \eq{deltaW} and similarly the field $\phi(\bfx,t)$,  we find equations of motion for perturbations $\delta\phi(\bfx,t)$ and $\delta \bfw(\bfx,t)$ from \eqs{EoM-phi}{EoM-W}. Keeping only the first order terms and switching to the Fourier space they become
\begin{eqnarray}
& &\ddot{\delta\phi}+3H\dot{\delta\phi}+\left(\mathbf k / a \right)^2\delta\phi+V''\delta\phi = \nonumber\\
& & \quad\quad= 2\frac{f'}{f}\left[\dot{\mathbf{W}}+\left(H-\frac{\dot{f}}{f}\right)\mathbf{W}\right]\left[\delta\dot{\mathbf{W}}+\left(H-\frac{\dot{f}}{f}\right)\delta\mathbf{W}-\delta\left(\frac{\dot{f}}{f}\right)\mathbf{W}\right], \dlabel{prtb-phi}\\
& &\delta\ddot{\mathbf{W}}+3H\delta\dot{\mathbf{W}}+\left(2H^{2}-H\frac{\dot{f}}{f}-\frac{\ddot{f}}{f}\right)\delta\mathbf{W}-a^{-2}\nabla^{2}\delta\mathbf{W}= \nonumber\\
& & \quad\quad=\left[H\delta\left(\frac{\dot{f}}{f}\right)+\delta\left(\frac{\ddot{f}}{f}\right)+\frac{f'}{f}\left(\frac{\mathbf k}{a}\right)^2\delta\phi\right]\mathbf{W}.\dlabel{prtb-W}
\end{eqnarray}
For exponentially varying gauge kinetic function $f$ in \eq{fofphi} the above expressions become
\begin{eqnarray}
& &\ddot{\delta\phi}+3H\dot{\delta\phi}+\left(\mathbf k / a \right)^2\delta\phi+V''\delta\phi=\nonumber \\
& &\quad\quad=\frac{2\alpha}{\sqrt{2\epsilon}\mpl}\left[\dot{\mathbf{W}}+H\left(1-\alpha\right)\mathbf{W}\right]\left[\delta\dot{\mathbf{W}}+H\left(1-\alpha\right)\delta\mathbf{W}+\alpha\frac{H}{\dot{\phi}}\mathbf{W}\dot{\delta\phi}\right],
\dlabel{a9}  \\
& &\delta\ddot{\mathbf{W}}+3H\delta\dot{\mathbf{W}}+\mu^{2}\delta\mathbf{W}+\left(\frac{k}{a}\right)^{2}\delta\mathbf{W}= \nonumber \\
& &\quad\quad=\frac{\alpha\mathbf{W}}{\sqrt{2\epsilon}\mpl}\left[\ddot{\delta\phi}+H\left(1+2\alpha\right)\dot{\delta\phi}+\left(\frac{k}{a}\right)^{2}\delta\phi\right], \dlabel{a10}
\end{eqnarray}
where $\mu^2= (2+\alpha)(1-\alpha)H^2$ and $\alpha \simeq -2$.

The energy density of the vector field in \eq{action} is given by \cite{fF2curv} $\rho_{B}(\bfx,t)=-f^{2}F_{\mu\nu}F^{\mu\nu}/4$. From this it is easy to see that the background value of $\rho_{B}(\bfx,t)$ is given by
\be
\rho_{B}(t)=\frac{1}{2}f^{2}\left(\frac{\dot{B}}{a}\right)^{2}=\frac{1}{2}\left[\dot{W}+\left(H-\frac{\dot{f}}{f}\right)W\right]^{2}\simeq\frac12 H^2 W^2.
\ee
The right hand side of 
 Eq.\eqref{hom-phi}
is negligible if $\rho_{B}$ satisfies \eq{rhowcon}. 
We now show that the same is true of the right hand sides of
\eqs{a9}{a10}.
At the epoch $k\sim aH$,  the terms on the left hand sides
are of order $H^3$ and \eq{rhowcon} ensures that the right hand sides
are indeed much smaller. At the epoch $aH/k =\exp(N_k(t))\gg 1$,
the first term of each left hand side is negligible.
The other two terms are of order $|\eta|\equiv |V''|
/3H^2$ for \eq{a9}  and of order $|\eta_W|\equiv |\mu^2|/3H^2$
for \eq{a10}.
\eq{rhowcon} ensures that the right hand side of \eq{a9} is negligible,
and it ensures that the right hand side of \eq{a10} is negligible
if also $|\eta_W|\gg 10\mfive$. But the latter condition is irrelevant, because
its violation makes the time-dependence of $W$ (coming then from
the right hand side) negligible.


\end{document}